\documentstyle[aps,preprint]{revtex}

\begin{document}
\draft \tightenlines \preprint{Alberta Thy 11-99;~gr-qc/9909002}

\title{Canonical Quantization of a Closed Euclidean Universe
with a Cosmological Constant}
\author{Sang Pyo Kim\footnote{Electronic address: sangkim@ks.kunsan.ac.kr}}
\address{Department of Physics \\ Kunsan National University
\\ Kunsan 573-701, Korea}

\date{\today}

\maketitle
\begin{abstract}
We study canonical quantization of a closed Euclidean universe
with a cosmological constant and a massless scalar field. The
closed Euclidean universe with an ordinary matter state can be
matched at a finite radius only with the closed Lorentzian
universe with the Wick-rotated exotic state. The exotic state
provides the Lorentzian universe with a potential barrier
extending from the cosmological singularity to the classical
turning point and corresponding to the Euclidean geometry through
the Wick-rotation, and avoid the singularity problem at the
matching boundary. We find analytically the approximate wave
functions for quantum creation of the Universe from the {\it
nothingness}. We prescribe the Hartle-Hawking's no-boundary wave
function, Linde's wave function and Vilenkin's tunneling wave
function. In particular, we find the wave function for the
Euclidean geometry, whose semiclassical solution is regular at the
matching boundary with the Lorentzian geometry but singular at the
cosmological singularity.
\end{abstract}

\section{Introduction}

Recently Hawking and Turok have proposed an open universe model
using a singular gravitational instanton for a closed Euclidean
Friedmann-Robertson-Walker (FRW) geometry coupled to a minimal
scalar field \cite{hawking-turok}. Further, they have used this
instanton to claim that the no-boundary proposal seems to be the
most favorable initial (boundary) condition for the wave function
of the Universe. This open universe model provoked again an
argument \cite{linde} and a counter-argument \cite{hawking-turok2}
among leading proposals for the initial condition of the wave
function of the Universe. Currently it is widely accepted that the
very early Universe emerged quantum mechanically from a Euclidean
regime into the Lorentzian regime of the present Universe. Leading
proposals for such wave functions are the Hartle-Hawking's
no-boundary wave function \cite{hartle-hawking}, Linde's wave
function \cite{linde2} and Vilenkin's tunneling wave function
\cite{vilenkin}. (For the differences among wave functions, see
Ref. \cite{vilenkin2}.)

Every Lorentzian FRW universe, regardless of its topology, with an
ordinary (positive energy density) matter state has a classical
allowed motion extending from the cosmological singularity to the
infinity. Hence, a Lorentzian FRW universe can be glued with a
Euclidean geometry only at the cosmological singularity of zero
radius. In that spirit the Vilenkin's instanton solution for a
closed Euclidean FRW universe with an ordinary state of a massless
scalar field matches an open Lorentzian FRW universe at the
cosmological singularity \cite{vilenkin3}. However, in spite of
some similar property to the Hawking-Turok's instanton, the
Vilenkin's instanton has a singularity at the matching boundary.
Even at the quantum cosmological context, the wave function of the
Lorentzian FRW universe shows an oscillatory behavior near the
cosmological singularity for every (closed or open or spatially
flat) topology provided that the massless scalar field has an
ordinary quantum state having a positive definite energy density
\cite{kim}. Thus, with an ordinary state of the scalar field a
Lorentzian FRW universe matches necessarily a Euclidean one at the
cosmological singularity.

In Refs. \cite{hawking-turok,linde,hawking-turok2,vilenkin3} an
open Lorentzian universe is matched with a closed Euclidean one.
In this paper we canonically quantize a closed FRW universe of
Lorentzian and Euclidean geometries with a massless scalar field
and a cosmological constant, find the wave functions that match
the Euclidean and Lorentzian geometries at a finite radius, and
study the matching condition for two geometries. We further
propose that an ordinary matter state in the Euclidean geometry
and the Wick-rotated exotic matter state in the Lorentzian
geometry lead to the genuine quantum creation of the Universe and
avoid any possible singularity problem at the matching boundary.
Indeed, it is only with the exotic quantum state of the scalar
field that the closed Lorentzian universe is matched at a finite
radius with the closed Euclidean universe. Therefore, quantum
creation of the Universe seems to require a mechanism to generate
a classically forbidden regime including the cosmological
singularity in the Lorentzian geometry.

One may naively adopt the quantum mechanical interpretation that
the classically forbidden regime under a potential barrier can be
interpreted as a tunneling regime with an imaginary time. In the
cosmological context this implies that a classically forbidden
regime of the Lorentzian geometry should correspond to a
classically allowed regime of the Euclidean geometry. Recently, by
using the Wick-rotation between the classically forbidden regime
of the Lorentzian geometry and the classically allowed regime of
the Euclidean geometry, we have found a rule to transform wave
functions which are well-defined in the Euclidean geometry into
those in the Lorentzian geometry \cite{kim1}. To treat properly
quantum creation of the Universe one needs to quantize not only
the Lorentzian geometry but also the Euclidean geometry and to
match wave functions of both geometries.

The organization of this paper is as follows. In Sec. II we study
the quantum FRW cosmological model with a cosmological constant
and a massless scalar field. It is found that the Wheeler-DeWitt
(WDW) equation has the correct wave function for quantum creation
of the Universe only when the scalar field has an exotic quantum
state. In Sec. III by using the approximate wave function found in
Sec. II we prescribe the Hartle-Hawking's no-boundary wave
function, Linde's wave function and Vilenkin's tunneling wave
function. In Sec. IV we study the same model in the Euclidean
geometry and find the wave function of the Euclidean geometry
corresponding to the classically forbidden regime of the
Lorentzian geometry. In Sec. V we derive the semiclassical
Einstein equation and find the asymptotic solution near the
matching boundary and the cosmological singularity. The
semiclassical solution from the wave function of the Euclidean
geometry is regular at the matching boundary with the Lorentzian
geometry but still singular at the cosmological singularity.

\section{A Model for Quantum Creation of Universe}

We consider the quantum cosmological model for the FRW universe
with a cosmological constant and a massless scalar field, which is
described by the action
\begin{equation}
I =  \frac{m_P^2}{16 \pi} \int d^4x \sqrt{-g}  \Bigl[R - 2 \Lambda
\Bigr] + \frac{m_P^2}{8 \pi} \int d^3x \sqrt{h} K -
\frac{1}{2}\int d^4x \sqrt{-g} g^{\mu \nu}
\partial_{\mu} \phi \partial_{\nu} \phi, \label{action}
\end{equation}
where $m_P = \frac{1}{\sqrt{G}}$ is the Planck mass, and $\Lambda$
and $\phi$ denote the cosmological constant and the scalar field,
respectively. The surface term for the gravity is introduced to
yield the correct equation of motion for the closed universe. In
the Arnowitt-Deser-Misner (ADM) formulation for the Lorentzian
geometry with the metric
\begin{equation}
ds_L^2 = - N^2 (t) dt^2 + a^2 (t) d \Omega^2_3,
\end{equation}
one obtains from the action (\ref{action}) the Hamiltonian
constraint
\begin{equation}
H_L = - \frac{1}{3 \pi m_P^2 a} \bigl(\pi_a \bigr)^2 - \frac{3 \pi
m_P^2}{4} \Bigl(a - \frac{\Lambda}{3} a^3\Bigr) + \frac{1}{4\pi^2
a^3}\bigl(\pi_{\phi}^L \bigr)^2 = 0, \label{L-con}
\end{equation}
where $\pi_a = - \frac{3 \pi m_P^2 a}{2N} \frac{\partial
a}{\partial t}$ and $\pi_{\phi}^L = \frac{2 \pi^2a^3}{N}
\frac{\partial \phi}{\partial t}$ are canonical momenta. According
to the Dirac quantization method, the Hamiltonian constraint
(\ref{L-con}) leads to the Wheeler-DeWitt (WDW) equation in the
Lorentzian geometry
\begin{equation}
\Biggl[- \frac{\hbar^2}{2m_P^2} \frac{\partial^2}{\partial a^2} +
\frac{9 \pi^2 m_P^2}{8} \Bigl(a^2 - \frac{\Lambda}{3}a^4 \Bigr) +
\frac{3 \hbar^2}{8 \pi a^2} \frac{\partial ^2}{\partial \phi^2}
\Biggr] \Psi_L (a, \phi) = 0. \label{L-wdw}
\end{equation}
When Eq. (\ref{L-wdw}) is regarded as a two-dimensional
zero-energy Schr\"{o}dinger equation, the standard interpretation
and techniques of quantum mechanics can be employed.

We consider two kinds of quantum states for the scalar field: one
is an ordinary matter state and the other is an exotic matter
state. First, let us consider the ordinary quantum state given by
\begin{equation}
\Phi_{L,or.}^{(\pm)} = \frac{e^{\pm i \hbar p \phi}}{(2 \pi)^{1/2}},
\label{LI-qst}
\end{equation}
which has a positive energy density with respect to the scalar
field Hamiltonian
\begin{equation}
\hat{H}_L  \Phi_{L,or.}^{(\pm)} =  \Biggl[ - \frac{3 \hbar^2}{8
\pi a^2} \frac{\partial ^2}{\partial \phi^2} \Biggr]
\Phi_{L,or.}^{(\pm)} = + \frac{3 \hbar^2 p^2}{8 \pi a^2}
\Phi_{L,or.}^{(\pm)}.
\end{equation}
Since the Hamiltonian of the massless scalar field is positive
definite in Lorentzian spacetimes, the quantum state
(\ref{LI-qst}) represents an ordinary matter. By inserting Eq.
(\ref{LI-qst}) into Eq. (\ref{L-wdw}), one separates the
gravitational field equation
\begin{equation}
\Biggl[- \frac{\hbar^2}{2m_P^2} \frac{\partial^2}{\partial a^2} +
V_{L,or.} (a) \Biggr] \psi_{L,or.} (a) = 0, \label{L-grav}
\end{equation}
where
\begin{equation}
V_{L,or.} (a) =  \frac{9 \pi^2 m_P^2}{8} \Bigl(a^2 -
\frac{\Lambda}{3}a^4 \Bigr) - \frac{3 \hbar^2 p^2}{8 \pi a^2}
\end{equation}
is an effective potential for the gravitational field. Equation
(\ref{L-grav}), the zero-energy Schr\"{o}dinger equation, has two
classically allowed regimes dominated by either the scalar field
for small $a$ or the cosmological constant for large $a$, which
are separated by a potential barrier dominated by the scalar
curvature. Then with the ordinary quantum state the Universe
should tunnel quantum mechanically from the Planckian FRW regime,
which is also classically accessible, to a Lorentzian de Sitter
spacetime. In this strict sense, the WDW equation (\ref{L-wdw})
with the ordinary quantum state can not describe genuinely quantum
creation of the Universe from the {\it nothingness} because there
exits a Planckian Lorentzian universe that includes the
cosmological singularity.

Next, let us consider an exotic quantum state given by
\begin{equation}
\Phi_{L,ex.}^{(\mp)} = \frac{e^{\mp \hbar \kappa \phi}}{(2 \pi)^{1/2}},
\label{LII-qst}
\end{equation}
which has a negative energy density with respect to the
Hamiltonian
\begin{equation}
\hat{H}_L  \Phi_{L,ex.}^{(\mp)} = \Biggl[ - \frac{3 \hbar^2}{8 \pi
a^2} \frac{\partial ^2}{\partial \phi^2} \Biggr]
\Phi_{L,ex.}^{(\mp)} = -\frac{3 \hbar^2 \kappa^2}{8 \pi a^2}
\Phi_{L,ex.}^{(\mp)}.
\end{equation}
The quantum state (\ref{LII-qst}) represents an exotic matter in
the sense that it has an imaginary momentum eigenvalue. However,
Eq. (\ref{LII-qst}) diverges in one side as $\phi \rightarrow \pm
\infty$. To have a well-behaved wave function at $a = 0$, as
required by the Hartle-Hawking's no-boundary wave function, one
needs to match smoothly both branches of Eq. (\ref{LII-qst}). This
can be done by defining a new quantum state, which is symmetric
with respect to $\phi = 0$:
\begin{equation}
\Phi^{\epsilon}_{L, ex.} = \lim_{\epsilon \rightarrow 0_+}
\frac{1}{(2 \pi)^{1/2}}
\exp \Bigl[- \hbar \kappa \tanh (\frac{\phi}{\epsilon}) \phi \Bigr].
\label{LIII-qst}
\end{equation}
Then Eq. (\ref{LIII-qst}) is bounded by $e^{ - \hbar \kappa
|\phi|}$ as $\phi \rightarrow \pm \infty$ and has the same first
and second derivatives as Eq. (\ref{LII-qst}) when the limit is
taken under the condition $|\phi| > \epsilon$ as $\phi \rightarrow
0$. Now the exotic matter contributes a back-reaction with the
opposite sign to the effective potential of the gravitational
field equation
\begin{equation}
\Biggl[- \frac{\hbar^2}{2m_P^2} \frac{\partial^2}{\partial a^2} +
V_{L,ex.} (a) \Biggr] \psi_{L,ex.} (a) = 0, \label{LII-grav}
\end{equation}
where
\begin{equation}
V_{L,ex.} (a) =  \frac{9 \pi^2 m_P^2}{8} \Bigl(a^2 -
\frac{\Lambda}{3}a^4 \Bigr) + \frac{3 \hbar^2 \kappa^2}{8 \pi
a^2}. \label{eff-potII}
\end{equation}
The effective potential has a classical turning point
approximately given by $a_t \approx \sqrt{\frac{3}{\Lambda}}$. The
classically forbidden regime is dominated by both the scalar
curvature and the scalar field, and the classically allowed regime
is dominated by the cosmological constant. Note that even the
Planck scale regime belongs to the classically forbidden regime.
Therefore, the WDW equation (\ref{L-wdw}) can describe quantum
creation of the Universe from the {\it nothingness} only when the
scalar field has the exotic quantum state (\ref{LII-qst}) or
(\ref{LIII-qst}).

From now on throughout this paper, we confine our study to the
exotic state (\ref{LII-qst}) or (\ref{LIII-qst}) and find
approximately wave functions in the classically allowed $(a
> a_t)$ and classically forbidden $(a < a_t)$ regimes. The
asymptotic solution valid throughout the whole regime can be found
by using the Liouville-Green transformation \cite{kim2}. First, in
the classically allowed regime, Eq. (\ref{LII-grav}) can be
rewritten as
\begin{equation}
\Bigl[\frac{\partial^2}{\partial \eta^2} + \eta + \Delta (\eta)
\Bigr] \Psi_{L,ex.} (\eta) = 0, \label{gr eq1}
\end{equation}
where
\begin{eqnarray}
\eta^{3/2} &=& \frac{3}{2} \int^{a}_{a_t} da \sqrt{-
\frac{2m_P^2}{\hbar^2} V_{L,ex.} (a) }\nonumber\\ \Delta (\eta)
&=& - \frac{3}{4} \frac{\Bigl( \frac{d^2 \eta}{d a^2} \Bigr)^2 }{
\Bigl( \frac{d \eta}{d a} \Bigr)^4 } +  \frac{1}{2} \frac{ \Bigl(
\frac{d^3 \eta}{d a^3} \Bigr) }{\Bigl(\frac{d \eta}{d a}
\Bigr)^3}.\label{trans1}
\end{eqnarray}
In this regime the scalar field contributes insignificantly, so we
solve approximately Eq. (\ref{trans1}) to obtain
\begin{eqnarray}
\eta &=& \Bigl(\frac{9 \pi m_P^2}{4 \hbar \Lambda} \Bigr)^{2/3}
\Bigl( \frac{\Lambda}{3}a^2 -1 \Bigr), \nonumber\\ \Delta (\eta)
&=& - \frac{1}{16} \frac{1}{\Bigl[\eta + \Bigl(\frac{9 \pi
m_P^2}{4 \hbar \Lambda} \Bigr)^{2/3} \Bigr]^2}.
\end{eqnarray}
For $|\eta | >> 1$, $\Delta (\eta) = - \frac{1}{16 \eta^2}$, and
for $|\eta| << 1$, $\Delta (\eta) \simeq \Delta_0$, where
$\Delta_0 = - \frac{1}{16} \Bigl( \frac{4 \hbar \Lambda}{9 \pi
m_P^2}\Bigr)^{4/3}$. It is worthy to mention that $\Delta (\eta)$
approaches to a constant value even at the turning point. Then the
wave function has two branches
\begin{eqnarray}
\Psi_{L,ex.} (a, \phi) = C_{L} \Phi_{L, ex.} (\phi)
\left\{\begin{array}{l} Ai (- \eta - \delta (\eta))
\\ Bi (- \eta - \delta(\eta)) \end{array} \right\}. \label{L-waveII}
\end{eqnarray}
Here $\Phi_{L, ex.} (\phi)$ denotes either Eq. (\ref{LII-qst}) or
(\ref{LIII-qst}), depending on the behavior required at $\phi
\rightarrow \pm \infty$, and $Ai (-\eta)$ and $Bi (-\eta)$ are the
Airy functions \cite{abramowitz}, and $\delta (\eta) \simeq
\Delta_0$ for $|\eta| << 1$ and $\delta (\eta) \simeq -
\frac{1}{80 \eta^2}$ for $|\eta|
>> 1$. These wave functions are real and show an oscillatory
behavior.

The analytic wave function is also found approximately in the
classically forbidden regime. In this regime we change the
variable as
\begin{eqnarray}
\zeta^{3/2} &=& \frac{3}{2} \int_{a}^{a_t} da
\sqrt{\frac{2m_P^2}{\hbar^2} V_{L,ex.}(a)}, \nonumber\\ \Delta
(\zeta) &=& - \frac{3}{4} \frac{\Bigl( \frac{d^2 \zeta}{d a^2}
\Bigr)^2 }{ \Bigl( \frac{d \zeta}{d a} \Bigr)^4 } +  \frac{1}{2}
\frac{ \Bigl( \frac{d^3 \zeta}{d a^3} \Bigr) }{\Bigl(\frac{d
\zeta}{d a} \Bigr)^3}, \label{trans2}
\end{eqnarray}
and rewrite Eq. (\ref{LII-grav}) as
\begin{equation}
\Bigl[\frac{\partial^2}{\partial \zeta^2} - \zeta + \Delta (\zeta)
\Bigr] \Psi_{L,ex.} (\zeta) = 0. \label{gr eq2}
\end{equation}
We solve Eq. (\ref{trans2}) approximately to get
\begin{eqnarray}
\zeta &=& \Bigl(\frac{9 \pi m_P^2}{4 \hbar \Lambda} \Bigr)^{2/3}
\Bigl(1-  \frac{\Lambda}{3}a^2 \Bigr), \nonumber\\ \Delta (\zeta)
&=& - \frac{1}{16} \frac{1}{\Bigl[\zeta + \Bigl(\frac{9 \pi
m_P^2}{4 \hbar \Lambda} \Bigr)^{2/3} \Bigr]^2},
\end{eqnarray}
and find the wave function
\begin{eqnarray}
\Psi_{L,ex.} (a, \phi) = C_{L} \Phi_{L, ex.} (\phi)
 \left\{\begin{array}{l} Ai (\zeta -
\delta (\zeta) )
\\ Bi (\zeta - \delta (\zeta) ) \end{array} \right\}. \label{L-waveIII}
\end{eqnarray}

A few comments are in order. Equation (\ref{gr eq2}) is the
analytic continuation of Eq. (\ref{gr eq1}) through $\zeta = -
\eta$. Moreover, the fact that $\delta (\eta)$ or $\delta (\zeta)$
shifts only by a phase even at the turning point $\eta = 0$ or
$\zeta = 0$ justifies the validity of the Liouville-Green
transformation as a good approximation even near the classical
turning point. For large argument $\zeta$, the Airy function $Ai
(\zeta)$ is exponentially decreasing, but $Bi(\zeta)$ is
exponentially increasing. We also point out the fact that the
classically forbidden regime corresponds to a classically allowed
regime of the Euclidean geometry. In the Euclidean geometry one
has $\pi^E_a = i \pi^L_a$ and $\pi^E_{\phi} = i \pi^L_{\phi}$, so
the scalar field with the exotic quantum state, Eqs.
(\ref{LII-qst}) or (\ref{LIII-qst}), has a real momentum in the
Euclidean geometry and becomes an ordinary matter \cite{kim1}.
Hence gluing the Euclidean FRW universe with the Lorentzian one
can be explained naturally as quantum creation.

\section{Prescription for the Wave Functions}

Using the approximate wave function in Sec. II, we may prescribe
leading proposals, the Hartle-Hawking's no-boundary wave function,
Linde's wave function and Vilenkin's tunneling wave function. We
first consider the Hartle-Hawking's no-boundary wave function. In
the classically forbidden regime $a < a_t \approx
\sqrt{\frac{3}{\Lambda}}$, the no-boundary wave function, which is
bounded for all $\phi$ at $a = 0$, is given by the wave function
\begin{equation}
\Psi_{HH} (a, \phi) = C_{L} \Phi^{\epsilon}_{L, ex.} (\phi)
 Ai (\zeta - \delta (\zeta) ). \label{hh}
\end{equation}
From the asymptotic expansion of $Ai (\zeta)$ for large $\zeta$
\cite{abramowitz}, we obtain approximately the wave function
\begin{equation}
\Psi_{HH} (a, \phi) = C_{L} \Phi^{\epsilon}_{L, ex.} (\phi)
\Bigl(\frac{1}{1 - \frac{\Lambda}{3} a^2}
\Bigr)^{1/4} e^{- S_g (a)},
\end{equation}
where
\begin{equation}
S_g (a) = \int^{a_t}_a da \sqrt{\frac{2m_P^2}{\hbar^2} V_{L, ex.}
(a)}. \label{inst}
\end{equation}
$S_g$ is the gravitational instanton under the potential barrier.
Near the cosmological singularity, $\zeta$ and $S_g$ become
sufficiently large, so the no-boundary wave function is regular at
the cosmological singularity. In the classically allowed region,
the no-boundary wave function can be obtained by analytically
continuing $\zeta$ of Eq. (\ref{trans2}) beyond the classical
turning point, which is nothing but $\zeta = - \eta$. The
corresponding wave function is one branch of solutions of Eq.
(\ref{L-waveII})
\begin{equation}
\Psi_{HH} (a, \phi) = C_{L} \Phi^{\epsilon}_{L, ex.} (\phi) Ai (-
\eta - \delta (\eta) ), \label{hh-2}
\end{equation}
and has the asymptotic form
\begin{equation}
\Psi_{HH} (a, \phi) = C_{L} \Phi^{\epsilon}_{L, ex.} (\phi)
 \Bigl(\frac{1}{1 - \frac{\Lambda}{3} a^2}
\Bigr)^{1/4} \sin \Bigl(\int^{a}_{a_t} da \sqrt{-
\frac{2m_P^2}{\hbar^2} V_{L,ex.} (a)} + \frac{\pi}{4} \Bigr).
\end{equation}

Second, the Linde's wave function in the classically forbidden
regime is prescribed by
\begin{equation}
\Psi_{Lin} (a, \phi) = C_{L} \Phi_{L, ex.} (\phi) Bi (\zeta - \delta (\zeta)). \label{linde}
\end{equation}
From the asymptotic form
\begin{equation}
\Psi_{Lin} (a, \phi) = C_{L} \Phi_{L, ex.} (\phi) \Bigl(\frac{1}{1 - \frac{\Lambda}{3} a^2}
\Bigr)^{1/4} e^{+ S_g (a)},
\end{equation}
one sees that the Linde's wave function is exponentially large at
the cosmological singularity. In the classically allowed region
the Linde's wave function is given by the other branch of
solutions of Eq. (\ref{L-waveII})
\begin{equation}
\Psi_{Lin} (a, \phi) = C_{L} \Phi_{L, ex.} (\phi) Bi (- \eta - \delta (\eta)), \label{linde-2}
\end{equation}
and has the asymptotic form
\begin{equation}
\Psi_{Lin} (a, \phi) = C_{L} \Phi_{L, ex.} (\phi) \Bigl(\frac{1}{1
- \frac{\Lambda}{3} a^2} \Bigr)^{1/4} \cos \Bigl(\int^{a}_{a_t} da
\sqrt{- \frac{2m_P^2}{\hbar^2} V_{L,ex.} (a)} + \frac{\pi}{4}
\Bigr).
\end{equation}

Finally, the Vilenkin's tunneling wave function in the classically
forbidden regime is prescribed by
\begin{equation}
\Psi_{V} (a, \phi) = C_{L} \Phi_{L, ex.} (\phi) \Bigl[ Ai (\zeta - \delta (\zeta)) - i Bi
(\zeta - \delta (\zeta)) \Bigr]. \label{vilenkin}
\end{equation}
The tunneling wave function has the asymptotic form
\begin{equation}
\Psi_{V} (a, \phi) = C_{L}\Phi_{L, ex.} (\phi) \Bigl(\frac{1}{1 - \frac{\Lambda}{3} a^2}
\Bigr)^{1/4}  \Bigl[ \frac{1}{2} e^{- S_g (a)} +  e^{+ S_g (a)}
\Bigr].
\end{equation}
It is a linear superposition of the exponentially growing and
damping branches of wave functions. Similarly, in the classical
allowed regime the tunneling wave function is given by the linear
superposition of two branches of solutions of Eq. (\ref{L-waveII})
\begin{equation}
\Psi_{V} (a, \phi) = C_{L} \Phi_{L, ex.} (\phi) \Bigl[ Ai (-\eta -
\delta (\eta)) - i Bi (-\eta - \delta (\eta)) \Bigr].
\label{vilenkin2}
\end{equation}
In another form it describes obviously the expanding Universe
\begin{equation}
\Psi_{L} (a, \phi) = C_{L} \Phi_{L, ex.} (\phi) \eta^{1/2}
H^{(1)}_{1/3} (\frac{2}{3}
\eta^{3/2}), \label{vilenkin3}
\end{equation}
where $H^{(1)}_{1/3}$ is the Hankel function of the first kind.

\section{Canonical Quantization of Euclidean Geometry}

We now turn to canonical quantization of the Euclidean geometry.
In the ADM formulation for the Euclidean geometry with the
$O(4)$-symmetric metric
\begin{equation}
ds_E^2  =  N^2 (\tau) d\tau^2 + b^2 (\tau) d \Omega_3^2,
\label{e-metric}
\end{equation}
the action (\ref{action}) leads to the Hamiltonian constraint
\begin{equation}
H_E =  \frac{1}{3 \pi m_P^2 b} \bigl( \pi_b \bigr)^2 - \frac{3 \pi
m_P^2 }{4} \Bigl(b - \frac{\Lambda}{3}b^3 \Bigr) - \frac{1}{4\pi^2
b^3}\bigl(\pi_{\phi}^E \bigr)^2 = 0,\label{E-con}
\end{equation}
where $\pi_b = - \frac{3 \pi m_P^2 b}{2N} \frac{\partial b}{
\partial \tau}$ and $\pi_{\phi}^E = \frac{2 \pi^2 b^3}{N} \frac{\partial
\phi}{\partial \tau}$. Canonical quantization of the closed
Euclidean universe proceeds via the Dirac method by imposing the
Hamiltonian constraint (\ref{E-con}) as the WDW equation
\begin{equation}
\Biggl[- \frac{\hbar^2}{2m_P^2} \frac{\partial^2}{\partial b^2} -
\frac{9 \pi^2 m_P^2}{8} \Bigl(b^2 - \frac{\Lambda}{3}b^4 \Bigr) +
\frac{3 \hbar^2}{8 \pi b^2} \frac{\partial ^2}{\partial \phi^2}
\Biggr] \Psi_E (b, \phi) = 0. \label{E-wdw}
\end{equation}
It is worthy to notice that the closed Euclidean universe very
close to $b = 0$ or without $\Lambda$ has the same equation
(\ref{E-wdw}) as the open Lorentzian FRW universe with the similar
restriction. In this sense, for the same type of matter the closed
Euclidean universe is just an extension of the open Lorentzian
universe beyond the cosmological singularity
\cite{hawking-turok,linde,hawking-turok2,vilenkin3}. However, this
extension differs from the analytical continuation of the
Lorentzian geometry through the Wick-rotation, on which we are
mostly concerned in this paper.

Though the energy for the gravity-matter system vanishes, both the
scalar and gravitational fields can oscillate throughout the
evolution of the universe, because the curvature of the Euclidean
universe plays the role of a negative potential for the
gravitational field (scale factor). We thus find an ordinary
quantum state given by
\begin{equation}
\Phi_E^{(\pm)} =  \frac{e^{\pm i k \phi}}{(2\pi)^{1/2}},
\end{equation}
which is a positive energy eigenstate of the scalar field
Hamiltonian
\begin{equation}
\hat{H}_E (\phi, \tau) = - \frac{3 \hbar^2}{8 \pi b^2}
\frac{\partial ^2}{\partial \phi^2}. \label{E-qst}
\end{equation}
The WDW equation ({\ref{E-wdw}) is then separated into the
gravitational field equation
\begin{equation}
\Biggl[- \frac{\hbar^2}{2m_P^2} \frac{\partial^2}{\partial b^2} +
V_E (b) \Biggr] \Psi_E (b) = 0, \label{E-grav}
\end{equation}
where
\begin{equation}
V_E (b) =  - \frac{9 \pi^2 m_P^2}{8} \Bigl(b^2 -
\frac{\Lambda}{3}b^4 \Bigr) - \frac{3 \hbar^2 k^2}{8 \pi b^2}
\label{e-pot}
\end{equation}
is the effective potential for the gravitational field. Again it
is worth pointing out that the gravitational field equation
(\ref{E-grav}) of the Euclidean geometry is exactly the inverted
motion (\ref{LII-grav}) of the Lorentzian geometry, which enables
us to interpret the Lorentzian regime under the potential barrier
as a Euclidean regime \cite{kim1}. This is quite analogous to the
instanton motion of a classical field in Minkowski spacetimes. In
the Euclidean geometry the scalar field contributes a
back-reaction to the effective potential, which provides a
classical motion near $b = 0$, whereas the cosmological constant
acts as a potential barrier to the Euclidean geometry, which is in
contrast with the Lorentzian geometry. The classically allowed
regime is largely determined by the cosmological constant and
extends to the turning point $b_t \approx
\sqrt{\frac{3}{\Lambda}}$.

As in the case of the Lorentzian geometry, we may find
analytically the approximate wave function in the classically
forbidden $(b
> b_t \approx \sqrt{\frac{3}{\Lambda}})$ and
allowed $(b_t < b)$ regimes. However, we are mostly interested in
the wave function of the Euclidean geometry in the classically
allowed regime, which is to be matched with the Lorentzian wave
function. We again make use of the Liouville-Green transformation
\cite{kim2}, which now reads
\begin{equation}
\Bigl[\frac{\partial^2}{\partial \chi^2} + \chi + \Delta (\chi)
\Bigr] \Psi_E (\chi) = 0,
\end{equation}
where
\begin{eqnarray}
\chi^{3/2} &=& \frac{3}{2} \int^{b_t}_b db \sqrt{-
\frac{2m_P^2}{\hbar^2} V_E (b)}, \nonumber\\ \Delta (\chi) &=& -
\frac{3}{4} \frac{\Bigl( \frac{d^2 \chi}{d b^2} \Bigr)^2 }{ \Bigl(
\frac{d \chi}{d b} \Bigr)^4 } +  \frac{1}{2} \frac{ \Bigl(
\frac{d^3 \chi}{d b^3} \Bigr) }{\Bigl(\frac{d \chi}{d b}
\Bigr)^3}. \label{trans3}
\end{eqnarray}
In the regime $b_t > b \gg 0$, the contribution from the scalar
field is negligible, so we approximately solve Eq. (\ref{trans3})
to get
\begin{eqnarray}
\chi &=& \Bigl(\frac{9 \pi m_P^2}{4 \hbar \Lambda} \Bigr)^{2/3}
\Bigl( 1 - \frac{\Lambda}{3}b^2 \Bigr), \nonumber\\ \Delta (\chi)
&=& - \frac{1}{16} \frac{1}{\Bigl[\chi - \Bigl(\frac{9 \pi
m_P^2}{4 \hbar \Lambda} \Bigr)^{2/3} \Bigr]^4}.
\end{eqnarray}
By noting that $\Delta(\chi)$ is small for large $\chi$ and
approaches a constant value even at the turning point $\chi = 0$,
we find the wave function
\begin{eqnarray}
\Psi_E^{(\pm)} (b, \phi) = C_E \frac{e^{\pm i k
\phi}}{(2\pi)^{1/2}} \left\{\begin{array}{l} Ai (- \chi - \delta
(\chi))
\\ Bi (- \chi - \delta (\chi)) \end{array} \right\}. \label{E-waveII}
\end{eqnarray}
Equations (\ref{L-waveII}) and (\ref{E-waveII}) are wave functions
in the classically allowed regimes of the Lorentzian and Euclidean
geometries, respectively. These wave functions are transformed
according to the rule in Ref. \cite{kim2}, from which the wave
function in the classically forbidden regime of the Lorentzian
geometry is obtained by the correspondence $\chi = \eta = -
\zeta$, and $k = i \kappa$ from $\pi_{E \phi} = i \pi_{\phi}$.

\section{Semiclassical Approach to Gravitational Instanton}

To exploit the meaning of the wave function in Sec. IV we adopt
the semiclassical (quantum) gravity approach \cite{kiefer} and
compare the result with the classical gravity. In the
semiclassical (quantum) gravity approach one sets the wave
function to have the form
\begin{equation}
\Psi_E^{(\pm)} (b) = F (b) \exp \Bigl[\pm \frac{i}{\hbar} S_E (b)
\Bigr].
\end{equation}
And one introduces the cosmological (WKB) time
\begin{equation}
\frac{\partial}{\partial \tau} = \mp \frac{2}{3 \pi m_P^2 b}
\Bigl( \frac{\partial S_E}{\partial b}\Bigr)
\frac{\partial}{\partial b}
\end{equation}
along a classical trajectory where the oscillating wave function
is peaked. Then along the classical trajectory one obtains the
following semiclassical Einstein equation for the Euclidean
geometry
\begin{equation}
\Biggl(\frac{\partial b /\partial \tau}{b} \Biggr)^2 -
\frac{1}{b^2}  + \frac{\Lambda}{3} = \frac{\hbar^2 k^2}{3 \pi^3
m_P^2 b^6}, \label{E-sem ein}
\end{equation}
and the time-dependent Schr\"{o}dinger equation for the scalar
field
\begin{equation}
i \hbar \frac{\partial}{\partial \tau} \Phi_E (\phi, \tau) =
\hat{H}_E  \Phi_E (\phi, \tau).
\end{equation}

Let us now compare the semiclassical result with the classical one
by considering the $O(4)$-symmetric closed Euclidean FRW universe,
the metric of which is given by setting $N(\tau) = 1$ in Eq.
(\ref{e-metric}):
\begin{equation}
ds_E^2  = d\tau^2 + b^2_c (\tau) d \Omega^2_3,
\end{equation}
The classical equations of motion consist of the Einstein equation
\begin{equation}
\Bigl( \frac{\partial b_c / \partial \tau}{b_c}\Bigr)^2 -
\frac{1}{b^2_c}  + \frac{\Lambda}{3} = \frac{4 \pi}{3m_P^2}
\dot{\phi}_c^2, \label{cl ein eq}
\end{equation}
and the scalar field equation
\begin{equation}
\frac{\partial^2 \phi_c}{\partial \tau^2} + 3 \frac{\partial b_c/
\partial \tau }{b_c} \frac{\partial \phi_c}{\partial \tau} = 0. \label{cl sc eq}
\end{equation}
It should be noted that provided $\hbar k = 2 \pi^2 p$, where $p =
b^3_c \dot{\phi}_c$ is an integration constant of Eq. (\ref{cl sc
eq}), the semiclassical Einstein equation (\ref{E-sem ein}) is the
same as the classical Einstein equation (\ref{cl ein eq}).
Equation (\ref{cl ein eq}) is nothing but the classical Einstein
equation for the open universe in the Lorentzian geometry. Eq.
(\ref{E-sem ein}) has an asymptotic solution
\begin{eqnarray}
b (\tau) \approx \cases{b (\tau_1) \Bigl(\frac{\tau -
\tau_0}{\tau_1 - \tau_0} \Bigr)^{1/3}, &$ \Bigl(\frac{\hbar^2
k^2}{3 \pi^3 m_P^2}\Bigr)^{1/4} > b \geq 0$, \cr b (\tau_2)
\Bigl[1 - \tanh^2 \Bigl(\frac{3 \pi m_P}{2} \sqrt{\frac{1}{2}}
(\tau - \tau_2) \Bigr) \Bigr]^{1/2}, &$ \sqrt{\frac{3}{\Lambda}}
\gg b
> \Bigl(\frac{\hbar^2 k^2}{3 \pi^3 m_P^2}\Bigr)^{1/4}$, \cr
\sqrt{\frac{3}{\Lambda}} \cosh \Bigl(\sqrt{\frac{\Lambda}{3}}
(\tau - \tau_t) \Bigr) , &$ \sqrt{\frac{3}{\Lambda}} \geq b \gg
\Bigl(\frac{\hbar^2 k^2}{3 \pi^3 m_P^2}\Bigr)^{1/4}$, \cr}
\end{eqnarray}
where $\tau_1 - \tau_0 = \frac{\pi m_P b_1^3}{\hbar \kappa}
\sqrt{\frac{\pi}{3}}$. We have thus obtained the semiclassical
solution for the closed Euclidean universe, which has one
prominent feature different from others. First, near the matching
boundary of the turning point $b_t \approx
\sqrt{\frac{3}{\Lambda}}$ the Euclidean universe expands according
to the exponential-law like a de Sitter spacetime, so the scalar
field is regular and given by
\begin{equation}
\phi (\tau) \approx \frac{\Lambda}{3} \Biggl[\frac{\sinh
\Bigl(\sqrt{\frac{\Lambda}{3}} (\tau - \tau_t) \Bigr)}{2\cosh^2
\Bigl(\sqrt{\frac{\Lambda}{3}} (\tau - \tau_t) \Bigr)} +
\frac{1}{2} \arctan \Bigl(\sinh \Bigl(\sqrt{\frac{\Lambda}{3}}
(\tau - \tau_t) \Bigr) \Bigr) \Biggr].
\end{equation}
However, near the cosmological singularity $b = 0$ all the
Euclidean universes expand, regardless of their topologies,
according to the power-law $b = \tau^{1/3}$ and the scalar field
diverges logarithmically, which is the characteristic feature of
the spatially flat universe with a stiff matter \cite{kim}. This
power-law is essentially the same result obtained by Hawking and
Turok for the Euclidean FRW universe with a minimal scalar field,
in which the kinetic energy dominates over the potential energy
\cite{hawking-turok}. In open universe models with ordinary
matters, due to the power-law of the scale factor, the scalar
field diverges logarithmically at the matching boundary of the
cosmological singularity, on which the closed Euclidean universe
is matched with an open Lorentzian universe
\cite{hawking-turok,linde,hawking-turok2,vilenkin3}. The
singularity problem may not be serious at the cosmological
singularity because of the finite total action but it can be
certainly at the matching boundary \cite{vilenkin3}.

\section{Conclusion}

In this paper we have studied canonical quantization of a closed
FRW universe with a massless scalar field and a cosmological
constant in both Lorentzian and Euclidean geometries. It is found
that only with an exotic state of the scalar field the wave
function of the Lorentzian geometry can describe the genuine
quantum creation of the Universe through a potential barrier from
the {\it nothingness}. Furthermore, the fact that the WDW equation
for the closed Euclidean geometry with an ordinary state is
exactly the inverted one for the closed Lorentzian geometry with
the Wick-rotated exotic state provides a matching condition for
two geometries at the turning point of a finite radius. This is
quite analogous to the instanton motion of a classical field in
Minkowski spacetimes. At the matching boundary of the finite
radius, the solutions to the Euclidean equations of motion are
well-behaved and the gravitational instanton and the scalar field
become regular. However, the possibility is excluded that an open
Lorentzian universe with an ordinary state may match the closed
Euclidean universe at a finite radius without causing any
singularity problem. Instead, the open Lorentzian universe with
the ordinary state can be matched with the closed Euclidean one
only at the cosmological singularity, but the Euclidean sector
leads necessarily to a singular instanton at the boundary.

The wave function is found approximately in the classically
forbidden and allowed regimes of the Lorentzian geometry, in terms
of which the Hartle-Hawking's no-boundary, Linde's wave function
and Vilenkin's tunneling wave function are prescribed. Since the
classically forbidden regime of the Lorentzian geometry
corresponds now to a classically allowed regime of the Euclidean
geometry, the wave function of the Lorentzian regime is the
Wick-rotation of that of the corresponding Euclidean regime. The
semiclassical limit of the wave function in the Euclidean geometry
recovers essentially the similar result obtained by Hawking and
Turok \cite{hawking-turok}, in which the kinetic energy of the
scalar field dominates over the potential energy, and in
particular, almost the same result obtained by Vilenkin for a
massless scalar field model \cite{vilenkin3}. However, our
semiclassical solution has a logarithmic divergence at the
cosmological singularity but becomes regular on the matching
boundary between the Euclidean and Lorentzian geometries. The
divergent solution does not cause any serious problem at the
cosmological singularity since the total action is finite.
Therefore, in canonical quantization of the closed Universe we are
able to avoid the singularity problem in evaluating the action for
the gravity-matter system.

The Wick-rotation between the Lorentzian and Euclidean geometries
allows us to find a spectrum of wave functions in the Lorentzian
geometry, which describe quantum creation of the Universe through
the potential barrier from the {\it nothingness}, from that of the
Euclidean geometry. In particular, the regular wave function at
the cosmological singularity may be interpreted as a quantum
wormhole \cite{hawking-page,kim2}, which exhibits an exponentially
damping behavior due to the scalar curvature of three geometry for
large size and a regular behavior for small size of the closed
Universe. Our model needs an exotic matter state in the Lorentzian
geometry, otherwise matter provides only a classically allowed
regime near the cosmological singularity, and also needs a
cosmological constant to let the Universe emerge from the
potential barrier that is dominated by the kinetic energy of the
scalar field and the scalar curvature of the three geometry.
However, the issue of the initial (boundary) condition of the wave
function for quantum creation of the Universe is not completely
settled, because canonical quantization leads to the WDW equation,
a relativistic wave functional equation for the gravity-matter
system. So at most one can obtain a spectrum of wave functions.

\acknowledgements
The author would like to thank D. N. Page for
many useful discussions and comments. He also would like to
appreciate the warm hospitality of APCTP and the Theoretical
Physics Institute, University of Alberta, where this paper was
completed. This work was supported by International Collaboration
Program of the Korea Research Foundation, 1997.

\end{document}